%Paper: hep-ph/9310332
%From: cho@theory3.caltech.edu (Peter Cho)
%Date: Thu, 21 Oct 93 10:08:24 PDT

% ---------------------------------------------------------------------
% This file uses the Harvmac macros and should be printed out in
% "big" format.
% ---------------------------------------------------------------------

\input harvmac

% The following command eliminates black boxes that appear when
% equations run over RHS margins:

\overfullrule=0pt

% Small capital subscripts

\def\C{{\scriptscriptstyle C}}
\def\D{{\scriptscriptstyle D}}

\def\L{{\scriptscriptstyle L}}
\def\M{{\scriptscriptstyle M}}

\def\R{{\scriptscriptstyle R}}
\def\S{{\scriptscriptstyle S}}

\def\U{{\scriptscriptstyle U}}

\def\Y{{\scriptscriptstyle Y}}

% Calligraphy letters

\def\CL{{\cal L}}

% Greek letters

\def\a{\alpha}
\def\b{\beta}

\def\e{\epsilon}
\def\g{\gamma}
\def\n{\eta}
\def\o{\sigma}

\def\u{\mu}
\def\v{\nu}
\def\z{\zeta}

%  Aliases

\def\221{SU(2)_\L \times SU(2)_\R \times U(1)}
\def\aEM{\alpha_{\scriptscriptstyle EM}}
\def\aS{\alpha_s}
\def\bar#1{\overline{#1}}
\def\ccdot{\hbox{\kern-.1em$\cdot$\kern-.1em}}
 % Slashed covariant deriv
		% \not{CP}
\def\dash{{\> \over \>}} 		% Hyphen in equations

\def\eff{{\rm\scriptscriptstyle eff}}
\def\EM{{\scriptscriptstyle EM}}
\def\fmcmb{f\bigl({\mc\over\mb}\bigr)}
\def\GeV{\>\, {\rm GeV}}
\def\GF{G_{\scriptscriptstyle F}}
\def\gfive{\gamma^5}
\def\gmcmb{g\bigl({\mc\over\mb}\bigr)}
\def\gone{g_{\scriptscriptstyle 1}}
\def\gtwoL{g_{\scriptscriptstyle 2L}}
\def\gtwoR{g_{\scriptscriptstyle 2R}}

\def\gthree{g_{\scriptscriptstyle 3}}
\def\gtap{\raise.3ex\hbox{$>$\kern-.75em\lower1ex\hbox{$\sim$}}}

\def\L{{\scriptscriptstyle L}}
\def\LR{{\scriptscriptstyle LR}}
\def\ltap{\raise.3ex\hbox{$<$\kern-.75em\lower1ex\hbox{$\sim$}}}
\def\mb{m_b}
\def\mc{m_c}
\def\mi{m_i}

\def\mt{m_t}

\def\MW1{M_{W_1}}

\def\R{{\scriptscriptstyle R}}

  % Macro for slashes thru vectors.
\def\sp{\>\>}
\def\TeV{\> {\rm TeV}}
\def\therefore{{\hbox{..}\kern-.43em \raise.5ex \hbox{.}}\>\>}
\def\twoone{SU(2)_\L \times U(1)_\Y}
\def\vL{v_\L}
\def\vR{v_\R}
\def\Wslash{W\hskip-0.8 em / \hskip+0.30 em}
\def\zg{\zeta_g}

% Fractions

\def\half{{1 \over 2}}

\def\sixth{{ 1\over 6}}

\def\twothirds{{2 \over 3}}

% Bold Greek letter macro:

% Style-sensitive Poor-Man's-Bold command, produces bold greek letters.
% Usage $ ... \pmb\gamma ... $
% Adapted from TeXbook p386 (\pmb) and p360 (\mathpallette)
\newdimen\pmboffset
\pmboffset 0.022em
\def\oldpmb#1{\setbox0=\hbox{#1}%
 \copy0\kern-\wd0
 \kern\pmboffset\raise 1.732\pmboffset\copy0\kern-\wd0
 \kern\pmboffset\box0}
\def\pmb#1{\mathchoice{\oldpmb{$\displaystyle#1$}}{\oldpmb{$\textstyle#1$}}
      {\oldpmb{$\scriptstyle#1$}}{\oldpmb{$\scriptscriptstyle#1$}}}

% Modified title definition to allow long titles to be broken up into three
% lines.

 %
% ----------------------------------------------------------------------
% References:
\nref\CLEO{CLEO Collaboration, R. Ammar {\it et al.}, Phys. Rev. Lett. {\bf 71}
 (1993) 674.}
\nref\QCDcorrections{
 B. Grinstein, R. Springer and M.B. Wise, Nucl. Phys. {\bf B339}  (1990)
  269\semi
 M. Misiak, Nucl. Phys. {\bf B393} (1993) 23\semi
 R. Grigjanis, P.J. O'Donnell, M. Sutherland and H. Navelet,
 Phys. Rept. {\bf 228} (1993) 93, and references therein.}
\nref\Ciuchini{M. Ciuchini, E. Franco, G. Martinelli, L. Reina and L.
 Silvestrini, preprint ROME 93/958 (hep-ph 9307364).}
\nref\twoHiggs{B. Grinstein and M.B. Wise, Phys. Lett. {\bf B201} (1988)
 274\semi
 W.-S. Hou and R.S. Willey, Phys. Lett. {\bf B202} (1988) 591\semi
 A.J. Buras, P. Krawczyk, M.E. Lautenbacher and C. Salazar, Nucl. Phys.
 {\bf B337} (1990) 284\semi
 J.L. Hewett, Preprint ANL-HEP-PR-92-110.}
\nref\SUSY{S. Bertolini, F. Borzumati, A. Masiero and G. Ridolfi, Nucl. Phys.
 {\bf B353} (1991) 591\semi
 R. Barbieri and G.F. Guidice, Phys. Lett. {\bf B309} (1993) 86\semi
 J.L. Lopez, D.V. Nanopoulos, G.T. Park, Preprint CTP-TAMU-16/93,
 hep-ph 9304277,}
\nref\Randall{L. Randall and R. Sundrum, MIT preprint MIT-CTP-2211 (1993).}
\nref\Mohapatra{R.N. Mohapatra and J.C. Pati, Phys. Rev. {\bf D11} (1975) 566;
 Phys. Rev. {\bf D11} (1975) 2558\semi
 R.N. Mohapatra and G. Senjanovic, Phys. Rev. {\bf D12} (1975) 1502.}
\nref\Senjanovic{G. Senjanovic, Nucl. Phys. {\bf B153} (1979) 334.}
\nref\Grimus{W. Grimus, Preprint UWTHPh-1993-10, and references
 therein.}
\nref\Cocolicchio{D. Cocolicchio, G. Costa, G.L. Fogli, J.H. Kim and A.
 Masiero, Phys. Rev. {\bf D40} (1989) 1477.}
\nref\Asatryan{G.M. Asatryan and A.N. Ioannisyan, Sov. J. Nucl. Phys. {\bf
 51} (1990) 858.}
\nref\LangackerI{P. Langacker in {\it CP Violation}, edited by C.
 Jarlskog (World Scientific, Singapore, 1989) 552.}
\nref\LangackerII{P. Langacker and S.U. Sankar, Phys. Rev. {\bf D40} (1989)
 1569.}
\nref\Beall{G. Beall, M. Bander and A. Soni, Phys. Rev. Lett. {\bf 48},
 (1982) 848.}
\nref\Abott{L. Abott, Nucl. Phys. {\bf B185} (1981) 189.}
\nref\Cho{P. Cho and B. Grinstein, Nucl. Phys. {\bf B365} (1991) 279.}
\nref\Inami{T. Inami and C.S. Lim, Prog. Theor. Phys. {\bf 65} (1981)
 297; (E) {\bf 65} (1981) 1772.}
\nref\Cabibbo{N. Cabibbo and L. Maiani, Phys. Lett. {\bf B79} (1978) 109.}
\nref\MisiakII{M. Misiak, Preprint TUM-T31-46/93, hep-ph 9309236.}
\nref\CDF{P. Tipton, Talk presented at the Lepton-Photon Conference,
 Cornell University, Aug. 1993.}

% ----------------------------------------------------------------------
% Figure captions:

\nfig\matchinggraphs{One-loop 1PI intermediate gauge and would-be
Goldstone boson graphs which contribute to the $b \to s \g$ matching
condition at the $W_1$ scale.   The circles at the ends of wavy
external propagators represent background photon fields.}
\nfig\uncorrectedzetaplot{Inclusive $\bar{B} \to X_s \g$ decay rate
normalized to the semileptonic $\bar{B} \to X_c e \bar{\v}_e$ rate plotted
as a function of the mixing angle $\z$ with $\mt=150 \GeV$ and $\a=0$.
The solid and dashed curves depict the QCD uncorrected results in $\221$ theory
and Standard Model respectively.}
\nfig\zetaplot{Inclusive $\bar{B} \to X_s \g$ decay rate normalized to
the semileptonic $\bar{B} \to X_c e \bar{\v}_e$ rate plotted as a
function of the mixing angle $\z$ with $\mt=150 \GeV$ and $\a=0$.
The solid and dashed curves depict the QCD corrected results in $\221$ theory
and Standard Model respectively.}
\nfig\mtplot{Inclusive $\bar{B} \to X_s \g$ decay rate normalized to
the semileptonic $\bar{B} \to X_c e \bar{\v}_e$ rate plotted as a
function of the top quark mass $\mt$ with $\z=0.0025$ and $\a=0$.
The solid and dashed curves depict the QCD corrected results in $\221$ theory
and Standard Model respectively.}
\nfig\alphaplot{Inclusive $\bar{B} \to X_s \g$ decay rate normalized to
the semileptonic $\bar{B} \to X_c e \bar{\v}_e$ rate plotted as a
function of the phase angle $\a$ with $\z=0.0025$ and $\mt=150 \GeV$.
The solid and dashed curves depict the QCD corrected results in $\221$ theory
and Standard Model respectively.}
%
% ----------------------------------------------------------------------
%
% Title page

\def\CITTitle#1#2#3{\nopagenumbers\abstractfont
\hsize=\hstitle\rightline{#1}
\vskip 0.6in\centerline{\titlefont #2} \centerline{\titlefont #3}
\abstractfont\vskip .5in\pageno=0}

\CITTitle{{\baselineskip=12pt plus 1pt minus 1pt
  \vbox{\hbox{CALT-68-1893}\hbox{DOE RESEARCH AND}\hbox{DEVELOPMENT REPORT}
  \hbox{TUM-T31-52/93}}}}
  {$b \to s \gamma$ Decay in $\221$ Extensions}{of the Standard Model}
\centerline{Peter Cho\footnote{$^\dagger$}{Work supported in part by
 by the U.S. Dept. of Energy under DOE Grant no. DE-FG03-92-ER40701
 and by an SSC Fellowship.}}
\centerline{Lauritsen Laboratory}
\centerline{California Institute of Technology}
\centerline{Pasadena, CA  91125}
\medskip\centerline{and}\medskip
\centerline{Miko\l aj Misiak\footnote{$^\ddagger$}{Work supported in
 part by the German Bundesministerium f\"ur Forschung und Technologie
 under contract 06 TM 732 and by the Polish Committee for Scientific
 Research.}}
\centerline{Physik-Department}
\centerline{Technische Universit\"at M\"unchen}
\centerline{85748 Garching, Germany}

\vskip .3in
\centerline{\bf Abstract}
\bigskip

	The rare radiative decay $b \to s \gamma$ is studied in
$SU(2)_\L \times SU(2)_\R \times U(1)$ extensions of the Standard Model.
Matching conditions for coefficients of operators appearing in the low
energy effective Hamiltonian for this process are derived, and QCD corrections
to these coefficients are analyzed.   The $b \to s \gamma$ decay rate is
then calculated and compared with the corresponding Standard Model result.
We find that observable deviations from Standard Model predictions can occur
in $SU(2)_\L \times SU(2)_\R \times U(1)$ theories for a reasonable range
of parameter values.

%\draft
\Date{10/93}
% ----------------------------------------------------------------------

\newsec{Introduction}

	The radiative weak decay $b \to s \g$ has been the subject of
significant experimental and theoretical study during the past several
years.  This rare transition has recently been observed for the first
time in the exclusive channel $\bar{B} \to K^* \g$ at CLEO \CLEO.  The
experimental bound on its inclusive rate has also been improved, and
better limits are expected to be set within the next few years.
On the theoretical side, $b \to s \g$ decay is of considerable interest
for several reasons.  Firstly, since this process involves third generation
fermions, it is sensitive to the heavy top quark and its rate grows with
increasing top mass.  Secondly, strong interaction corrections to this weak
radiative transition are known to be unusually large
\refs{\QCDcorrections,\Ciuchini}.  Two-loop diagrams that
generate the leading QCD corrections to this decay actually
dominate over the lowest order one-loop graphs.  But the most
exciting feature of this transition is its potential to reveal
departures from the Standard Model.  Since flavor structure remains
poorly understood, careful study of rare neutral flavor changing
processes offers one of the best prospects for glimpsing signs of new physics
in the near future.  The $b \to s \g$ transition thus provides a window onto
possible extensions of the Standard Model and has been investigated
in two-Higgs doublet models \twoHiggs, supersymmetric theories \SUSY\ and
extended technicolor scenarios \Randall.  Comparison of results from these
theories with experimental measurements places constraints upon
new physics which may lie beyond the Standard Model.

	In this paper, we examine $b \to s \g$ decay in another
well-known extension of the Standard Model.  Specifically, we consider
theories based upon the extended electroweak
gauge group $\221$.  Such models have been widely studied in the past
\refs{\Mohapatra{--}\Grimus}, and a number of phenomena such as
$K\dash\bar{K}$ mixing and neutrino masses have been
used to constrain their allowed parameter spaces.  The $b \to
s \g$ transition however has received relatively little attention within the
context of $\221$ theories.  We therefore will analyze this
important rare process in these models and compare the results with
those from the $\twoone$ theory.

	A previous study of the dominant gauge boson contributions to
$b \to s \g$ decay in $\221$ theories has been reported in
ref.~\Cocolicchio, while scalar contributions have been discussed in
ref.~\Asatryan.  Our work differs from and improves upon these earlier
findings in several important ways.  Firstly, we perform our
computations within the effective field theory framework which has become
standard in $b \to s \g$ investigations.  Comparison of
results between the $\twoone$ and $\221$ models is therefore
facilitated.  Use of effective field theory technology also allows us
to systematically incorporate QCD running effects which have
not been consistently treated before.  Secondly, we do not restrict
our analysis from the outset to models with manifest left-right
symmetry as previous authors have done.  Rather we allow for the more
general case of asymmetrical left and right handed sectors.  Finally,
our results differ both qualitatively and quantitatively from those
reported in the literature.  We therefore believe that our findings
provide several new insights into this problem.

	Our paper is organized as follows.  In section 2, we provide a
general review of $\221$ theories and present the particular model which
forms the basis of our $b \to s \g$ study.  In section 3, we derive a
low energy effective theory starting from the full $\221$ model, and we
calculate the coefficients of the leading nonrenormalizable operators
in its effective Hamiltonian which are relevant for $b \to s \g$ decay.
Strong interactions corrections are then discussed in section 4.
Finally, we evaluate the radiative decay rate for a
range of reasonable parameter values in the $\221$ theory and compare
our results with those from the Standard Model.

\newsec{The $\pmb{\221}$ Model}

	Theories based upon the electroweak gauge group $\221$
represent well-known extensions of the Standard Model.  Such theories
have been widely investigated both as simple generalizations of the
$\twoone$ model and as possible intermediate stages in grand unified
schemes like $SO(10)$.  One of the principle appeals of these models
is that they allow for parity to be restored as a symmetry of nature
at some energy scale above $250 \GeV$.  A discrete left-right
reflection has therefore commonly been imposed on most $\221$ models
which restricts their particle content and coupling constants.
The incorporation of parity represents however an
additional simplifying assumption which is not required by the
structure of the extended electroweak gauge group.  Moreover, left-right
symmetric theories are known to encounter difficulties if considered
in the context of grand unified models or cosmology
\refs{\LangackerI,\LangackerII}.  So more recent studies have focused
upon left-right asymmetric models.  In this article, we will
work within the framework of a general $\221$ model and not impose
left-right symmetry from the outset.

	To begin, we combine the color and electroweak sectors and start
with the extended gauge group $G = SU(3)_\C \times SU(2)_\L
\times SU(2)_\R \times U(1)$ which cascades down to the unbroken color
and electromagnetic subgroup $H=SU(3)_\C \times U(1)_\EM$ through the
following simple symmetry breaking pattern:
\eqn\pattern{\eqalign{
SU(3)_\C \> \times \> SU(2)_\L \> & \times \> SU(2)_\R \> \times \> U(1) \cr
T^a \qquad\qquad \> T^i_\L \quad \> & \qquad \> T^i_\R \> \qquad\quad S\cr
\gthree \qquad\qquad \> \gtwoL \quad \> & \qquad \> \gtwoR
\quad\qquad \gone \cr
& \downarrow \cr
SU(3)_\C \> \times \> SU&(2)_\L \> \times \> U(1)_\Y \cr
T^a \qquad\qquad & T^i_\L  \sp\quad Y/2=T^3_\R+S \cr
\gthree \>\qquad\qquad & \gtwoL \quad\qquad g' \cr
& \downarrow \cr
SU(3)_\C \sp & \times \sp U(1)_{\EM} \cr
T^a \sp\quad & \> \quad Q=T^3_\L+Y/2 \cr
\gthree \qquad & \quad\qquad e. \cr
}}
\bigskip\noindent
We have listed underneath each of the subgroup factors in this pattern
our nomenclature conventions for their associated generators and
coupling constants.  Our covariant derivative with respect to the
gauge group $G$ thus appears as
\eqn\covarderiv{D_\u = \partial_\u + i \gthree G^a_\u T^a
+ i \gtwoL W^i_{\L\u} T^i_\L + i \gtwoR W^i_{\R\u} T^i_\R + i \gone B_\u S.}

	We next display the fermion and scalar content of our model.
\foot{Throughout the remainder of this section, we adopt notation which
closely follows that established by Langacker and Sankar in
ref.~\LangackerII.}
Quarks and leptons
%
%\foot{Leptons will play no role in our $b \to s \g$ analysis and
%may be ignored.}
%
transform under $G$ as
\eqn\fermions{
\eqalign{q'_\L &= \pmatrix{u' \cr d' \cr}_\L \sim (3,2,1)^\sixth \cr
\ell'_\L &= \pmatrix{\v' \cr e' \cr}_\L \sim (1,2,1)^{-\half} \cr} \qquad
\eqalign{q'_\R &= \pmatrix{u' \cr d' \cr}_\R \sim (3,1,2)^\sixth \cr
\ell'_\R &= \pmatrix{\v' \cr e' \cr}_\R \sim (1,1,2)^{-\half} \cr}}
where the primes indicate that these fields are gauge rather than mass
eigenstates.  The fermions also carry a suppressed generation index which
ranges over three family values.  We introduce the scalar field
\eqn\Phiscalar{
\Phi = \pmatrix{\phi_1^0 & \phi_1^+ \cr \phi_2^- & \phi_2^0 \cr}
  \sim (1,2,\bar{2})^0}
which acquires the complex vacuum expectation value
\eqn\Phivev{\vev{\Phi} = \pmatrix{k & 0 \cr 0 & k' \cr}}
and generates fermion masses in the Yukawa sector.  After
diagonalization of the quark mass matrices, the primed gauge
eigenstate quark fields in \fermions\ are related to unprimed mass
eigenstate fields as
\eqn\masseigen{
   \eqalign{ u'_\L &= S_u u_\L \cr d'_\L &= S_d d_\L \cr} \qquad
   \eqalign{ u'_\R &= T_u u_\R \cr d'_\R &= T_d d_\R \cr}}
where $S_{u,d}$ and $T_{u,d}$ represent $3 \times 3$ unitary matrices
in family space.

	We need to include additional Higgs fields into our theory in
order to fully implement the symmetry breaking pattern specified in
\pattern.  There are a number of possibilities for how these scalars
may transform under $G$.  The rate for $b \to s \g$ decay in the
$\221$ model will not sensitively depend however upon the precise
structure of its scalar sector.  So we make the simplest choice and
introduce two doublet fields
\eqn\chiscalars{
\chi_\L = \pmatrix{\chi_\L^+ \cr \chi_\L^0} \sim (1,2,1)^\half \qquad
\chi_\R = \pmatrix{\chi_\R^+ \cr \chi_\R^0} \sim (1,1,2)^\half}
which acquire the real VEV's
\eqn\chivevs{
   \vev{\chi_\L} = \pmatrix{0 \cr \vL \cr} \quad {\rm and } \quad
   \vev{\chi_\R} = \pmatrix{0 \cr \vR \cr}.}
Although $\chi_\L$ is not essential for symmetry breaking
purposes, we incorporate it along with $\chi_R$ into the scalar sector
so that our model can be rendered left-right symmetric if desired.

	After the spontaneous symmetry breakdown $G \to H$, the kinetic
energy terms in the scalar Lagrangian
\eqn\Lscalar{\CL_{\rm scalar} = \Tr\bigl( D^\u \Phi^\dagger D_\u \Phi
\bigr) + D^\u \chi_\L^\dagger D_\u \chi_\L + D^\u \chi_\R^\dagger D_\u \chi_\R
- V(\Phi,\chi_\L,\chi_\R)}
generate the charged $W$ boson mass matrix
\eqn\Wmassmatrix{M_{W^\pm}^2 = \pmatrix{
\displaystyle{\gtwoL^2 \over 2} \bigl(\vL^2+|k|^2+|k'|^2 \bigr)
& - \gtwoL\gtwoR k^* k' \cr - \gtwoL\gtwoR k {k'}^* &
\displaystyle{\gtwoR^2\over 2} \bigl(\vR^2+|k|^2+|k'|^2 \bigr) \cr}
\equiv \pmatrix{M_\L^2 & M_{\L\R}^2 e^{i\a} \cr
                M_{\L\R}^2 e^{-i\a} & M_\R^2}}
where $\a$ represents the phase of $k^* k'$.  The eigenvalues
\eqn\Wmasses{\eqalign{M_1^2 &= M_\L^2 \cos^2\z + M_\R^2 \sin^2\z +
  M_{\L\R}^2 \sin 2\z \cr
  M_2^2 &= M_\L^2 \sin^2\z + M_\R^2 \cos^2\z -
  M_{\L\R}^2 \sin 2\z \cr}}
and eigenvectors
\eqn\Weigenstates{
\pmatrix{W_1^+ \cr W_2^+ \cr} = \pmatrix{\cos\z & e^{-i\a} \sin\z \cr
  -\sin\z & e^{-i\a} \cos\z} \pmatrix{W_\L^+ \cr W_\R^+ \cr}}
of this mass matrix correspond to the physical charged $W$ bosons in the
$\221$ theory.  The mass $M_2$ of the predominantly right handed $W_2$ as
well as the small $W_\L \dash W_\R$ mixing angle defined by
\eqn\tanzeta{\tan 2\z = - {2 M^2_{\L\R} \over M_\R^2 - M_\L^2}}
are restricted by a number of low energy phenomenological constraints.
Numerical estimates for bounds on these quantities in left-right
symmetric theories typically lie in the range \refs{\LangackerII,\Beall}
\eqn\numervals{M_2 > 1.4 \TeV \quad{\rm and}\quad |\z| < 0.0025.}
However in some corners of parameter space in particular $\221$ models,
$M_2$ masses as low as $300 \GeV$ or mixing angles as large as
$|\z| \approx 0.013$ are allowed.  So we will take the numbers in \numervals\
as reasonable estimates for these two important parameters but
consider ranges around these values as well.

	In order to maintain explicit gauge invariance in our Green's
functions, we will work in the background field version of
't~Hooft-Feynman gauge \Abott.  The gauge fixing Lagrangian in our model
schematically appears as
\eqn\LGF{\CL_{GF} = -{1 \over 2} \sum_a \bigl | \partial^\mu Q^a_\mu
- g_a f^{abc} {\overline Q}^b_\mu Q^{\mu c}
- i g_a \bigl( \phi^\dagger T^a \vev{\phi}
  - \vev{\phi}^\dagger T^a \phi \bigr) \bigr |^2. }
Here $Q^a_\mu$ represents a quantum gauge field for the gauge group
$G$, while ${\overline Q}^a_\mu$ stands for a classical background
field for the unbroken subgroup $H$.  As usual, the quadratic
$W_\u \partial^\u\phi$ cross terms that arise in the kinetic energy sector
of the scalar Lagrangian \Lscalar\ after spontaneous symmetry breaking are
canceled by identical terms in the gauge fixing Lagrangian.  The
expressions for the charged would-be Goldstone bosons corresponding to
the longitudinal components of the physical $W^+_{1,2}$ can simply be
read off from these quadratic cross terms:
\eqn\Goldstonebosons{\eqalign{
\pi_1^+ &= {\gtwoL \over \sqrt{2}} {\cos\z\over M_1} \Bigl[(- k'^* + z_1
k^*)\phi_1^+ + (k - z_1 k') \phi_2^+ - \vL \chi_L^+ - z_1 v_\R \chi_\R^+
\Bigr] \cr
\pi_2^+ &= {\gtwoR \over \sqrt{2}} {\cos\z\over M_2} e^{-i\a}
\Bigl[(k^* + z_2 {k'}^*)\phi_1^+ - (k' + z_2 k) \phi_2^+ + z_2 \vL \chi_L^+
- v_\R \chi_\R^+ \Bigr] \cr}}
where
\eqn\zfactors{z_1 = e^{-i\a} {\gtwoR\over\gtwoL} \tan\z \quad{\rm and}\quad
z_2 = e^{i\a} {\gtwoL\over\gtwoR} \tan\z.}
In addition, the following trilinear interactions between the
background photon field, physical $W_{1,2}$ bosons and would-be
Goldstone fields in the gauge fixing Lagrangian
\eqn\trilinearterms{\CL_{\rm GF} = \cdots + e M_1 \bar{A} W_1^+ \pi_1^-
+ e M_2 \bar{A} W_2^+ \pi_2^- + {\rm h.c.}}
are also canceled by terms in the Higgs kinetic energy sector.  This
extra cancellation results from our particular choice of
't~Hooft-Feynman background field gauge and will simplify our $b \to s
\g$ analysis.

	Having identified the Goldstone fields in
eqn.~\Goldstonebosons, we can readily derive their charged current
interactions.  It is important to note that the form of these
interactions is independent of our particular choice of scalar
representations in this model.  We display below the terms in the
charged current Lagrangian which are relevant for $b \to s \g$ decay:
\eqn\LCC{\eqalign{
\CL_{cc} = {1\over\sqrt{2}} \pmatrix{\bar{u}&\bar{c}&\bar{t}} \Bigl\{
& \Wslash_1^+ \Bigl[ - \gtwoL\cos\z V_\L P_- - \gtwoR\sin\z e^{i\a} V_\R P_+
\Bigr] \cr
+ & \Wslash_2^+ \Bigl[ \gtwoL\sin\z V_\L P_- - \gtwoR\cos\z e^{i\a} V_\R P_+
\Bigr] \cr
+ & {\pi_1^+ \over M_1} \Bigl[  \bigl(
\gtwoL\cos\z V_\L M_\D - \gtwoR e^{i\a}\sin\z M_\U V_\R \bigr) P_+ \cr
& \qquad - \bigl( \gtwoL\cos\z M_\U V_\L - \gtwoR e^{i\a} \sin\z V_\R M_\D
\bigr) P_- \Bigr]\cr
+ & {\pi_2^+ \over M_2} \Bigl[ - \bigl(
\gtwoL\sin\z V_\L M_\D+ \gtwoR e^{i\a} \cos\z M_\U V_\R \bigr) P_+ \cr
& \qquad +  \bigl(\gtwoL \sin\z M_\U V_\L + \gtwoR e^{i\a} \cos\z V_\R M_\D
\bigr) P_- \Bigr] \Bigr\} \pmatrix{d \cr s \cr b \cr}  + {\rm h.c.}
+ \cdots. \cr}}
In this expression, $P_\pm = (1 \pm \gfive)/2$ represent right and left
handed projection operators, $M_\U$ and $M_\D$ denote the diagonalized quark
mass matrices
\eqn\quarkmassmatrices{
M_\U = \pmatrix{m_u & 0 & 0 \cr
	        0 & m_c & 0 \cr
		0 & 0 & m_t \cr} \qquad
M_\D = \pmatrix{m_d & 0 & 0 \cr
		0 & m_s & 0 \cr
		0 & 0 & m_b \cr},}
and $V_\L = S_u^\dagger S_d$ and $V_\R = T_u^\dagger T_d$ are
the left and right handed analogs of the KM matrices in the $\221$
model.  In left-right symmetric theories, these KM matrices are related
as $|V_\L| = |V_\R|$ which clearly reduces the number of free parameters.

	Having set up our $\221$ model, we proceed to investigate $b \to s \g$
decay in the next section.

\newsec{The Effective Theory}

	The rare decay $b \to s \g$ is sensitive to new physics
above the electroweak scale $\vL$.  In most $\221$ extensions of the
Standard Model, the separation between $\vL$ and the scale $\vR$ where
the gauge group $G$ spontaneously breaks is quite large.  The
difference between the bottom quark and electroweak scales is also
large.  Therefore, this low energy radiative transition is especially well
suited for analysis within an effective field theory framework which
can take advantage of these large scale separations.

	The construction of the effective theory begins at $\mu=\vR$
in the $\221$ model.  Fields with masses of order this scale are
integrated out, but their virtual effects are incorporated into
nonrenormalizable operators whose coefficients are suppressed by
powers of $1/\vR$.  Since the lower bound on $\vR$ lies in the
multi-hundred GeV region, the contributions from $W_2^\pm$ and charged
physical scalars which naturally have $O(\vR)$ masses to $b \to s \g$
mediating operators are very small compared to those from $W_1^\pm$.
We therefore ignore such contributions and jump down to the $W_1$
scale where we simultaneously integrate out the top quark and charged
intermediate boson.  Our neglect of the splitting between the top and $W_1$
introduces an error.  However, its magnitude is known to be
approximately $10\%$ in the Standard Model \Cho, and we expect its
size in the $\221$ theory to be comparable.  We therefore will
tolerate this small uncertainty which could be systematically refined
if desired.

	The dominant one-loop contributions to $b \to s \g$ in the
$\221$ model come from the diagrams displayed in \matchinggraphs.  We
evaluate these graphs with their external propagators placed on-shell.
After performing an operator product expansion, we extract the leading
terms which match onto local magnetic moment operators.  Such terms
are generated only by the four 1PI diagrams shown in the figure.
Other 1PR graphs which arise at one-loop order do not match onto
magnetic moment operators and may therefore be ignored.

	It is sensible to make some simplifications
at this stage.  Firstly, since $\z$ is known to be quite small compared to
unity, we work only to $O(\z)$ and set $\cos\z \to 1$ and $\sin\z \to \z$.
Moreover, since $\z$ will always appear in combination with
$\gtwoR/\gtwoL$, we define $\zg=\gtwoR/\gtwoL \z$.  We also
neglect the mass of the strange quark relative to the bottom quark
mass.   The $b \to s \g$ amplitude is then given at the $W_1$ scale
by the {\it tree level} matrix element of the effective Hamiltonian
\eqn\HeffI{\eqalign{ H_\eff =
-{4 \GF\over\sqrt{2}} {e\mb\over 16\pi^2} \sum_{i=u,c,t}
{V^{is}_\L}^* V^{ib}_\L
\Bigl\{ &F(x_i) \, \bar{s} \sigma^{\u\v} P_+ b F_{\u\v} \cr
+ \zg {\mi\over\mb} &{\tilde F}(x_i) \, \bar{s} \sigma^{\u\v} \Bigl[
{V^{ib}_\R \over V^{ib}_\L} e^{i\a} P_+ + \Bigl({V^{is}_\R \over
V^{is}_\L} \Bigr)^* e^{-i\a} P_- \Bigr] b F_{\u\v} \Bigr\} \cr}}
where $x_i=(\mi/\MW1)^2$ and
\eqn\Ffunctions{\eqalign{
F(x) &= {x(7-5x-8x^2) \over 24(x-1)^3}
- {x^2(2-3x) \over 4(x-1)^4} \ln x \cr
{\tilde F}(x) &= {-20+31x-5x^2 \over 12 (x-1)^2}
+ {x(2-3x)\over 2(x-1)^3} \ln x. \cr}}

	 The first term inside the curly brackets in \HeffI\ is precisely
the same as in the Standard Model to which the $\221$ theory reduces in the
limit $\vR \to \infty$.  Its coefficient function $F$ is identical to the
analogous Standard Model function which was first calculated by
Inami and Lim \Inami.  On the other hand, the second term proportional
to ${\tilde F}$ represents a qualitatively new contribution to
$H_\eff$.  Since the physical $W_1$ boson in the $\221$ theory couples
to both left and right handed quarks, the one-loop diagrams in
\matchinggraphs\ can directly match onto odd dimension operators if the
intermediate charge $\twothirds$ quarks in these graphs undergo a
helicity flip.  The new terms arising from the
$\221$ theory are therefore proportional to $m_{i=u,c,t}$ rather than $\mb$.
Of course, the contribution coming from the virtual top quark is the
most important since it enhances the second term in eqn.~\HeffI\
relative to the first by $\mt/\mb$.  This contribution is further
enlarged by the ratio $r={\tilde F}(x_t)/F(x_t)$ which ranges over the
interval $7.7 \ge r \ge 3.5$ for $100 \GeV \le \mt \le 200 \GeV$.
So these two effects offset the suppression of the second
term in \HeffI\ by the small mixing angle $\zg$.  It is important to
note that no such enhancement occurs in the leading terms of
diagrams like those in \matchinggraphs\ with $W_1$ replaced by $W_2$.
So although $W_\L$-$W_\R$ mixing and $W_2$ exchange
are both $O(1/\vR^2)$ effects, the impact of the former
upon $b \to s \g$ decay in the $\221$ model is much more important than
the latter.

	Our matching results differ from those reported previously by
Cocolicchio {\it et al.} in ref.~\Cocolicchio.  In order to compare,
we have calculated all the necessary one-loop diagrams in ordinary as
well as background field 't~Hooft-Feynman gauge.  The expressions we
have obtained for the one-loop $W_1$ boson graphs in the ordinary
gauge are equivalent to the functions $F_{2\g a}^{\LR}(x)$ and $F_{2\g
b}^{\LR}(x)$ in eqn.~(17) of ref.~\Cocolicchio. However, the
contributions from the would-be Goldstone diagrams which we have
explicitly calculated disagree with the $F_{2\g c}^{\LR}(x)$ result of
Cocolicchio {\it et al}.  Their sum $F_{2\g a}^\LR+F_{2\g
b}^\LR+F_{2\g c}^\LR$ differs qualitatively and quantitatively from
our function ${\tilde F}(x)$.

	Having found the effective Hamiltonian expression in \HeffI, we
can easily take its tree level matrix element and compute the
$b \to s \g$ decay rate:
\eqn\bsphotrateI{
\Gamma(b \to s \g) = \displaystyle{{\GF^2 \mb^5 \over 32 \pi^4}} \aEM(\mb)
  \Bigl( |C|^2+|C'|^2 \Bigr)}
where
\eqn\Ccoeffs{\eqalign{
C &= \sum_{i=u,c,t} {V^{is}_\L}^* V^{ib}_\L \Bigl[ F(x_i) + \zg
{\mi\over\mb} {\tilde F}(x_i) {V^{ib}_\R \over V^{ib}_\L} e^{i\a} \Bigr]
\cr
C' &= \sum_{i=u,c,t} {V^{is}_\L}^* V^{ib}_\L \Bigl[
\zg {\mi\over\mb} {\tilde F}(x_i) \Bigl({V^{is}_\R \over V^{ib}_\L}
\Bigr)^*  e^{-i\a} \Bigr]. \cr}}
It is common practice to normalize this radiative partial width to the
semileptonic rate
\eqn\semirateI{
\Gamma(b \to c e \bar{\nu}_e) = {\GF^2 \mb^5 \over 192 \pi^3} |V^{cb}_\L|^2
\gmcmb}
where $g(\e)=1-8\e^2-24\e^4 \ln\e + 8\e^6-\e^8$ represents a phase
space factor \Cabibbo.  The sensitive dependence of
eqns.~\bsphotrateI\ and \semirateI\ upon the bottom quark mass and the
KM angles then cancels in their ratio
\eqn\rateratioI{R\equiv {\Gamma(b \to s \g) \over \Gamma(b \to c e \bar{\v}_e)}
\simeq {\Gamma(\bar{B} \to X_s \g) \over \Gamma(\bar{B} \to X_c e
\bar{\v}_e)}.}

	This ratio is plotted in \uncorrectedzetaplot\ as a function of the
mixing angle $\z$ with the top mass $\mt=150 \GeV$ and phase angle $\a=0$
held fixed, $\gtwoL$ set equal to $\gtwoR$, and all ratios of left and right
handed KM angles set equal to unity. In this left-right symmetric limit, the
up and charm quark contributions to the coefficients in \Ccoeffs\ are
completely negligible.  As can be seen in the figure, the QCD uncorrected
$b \to s \g$ rate in the $\221$ model is twice that in the Standard Model
for the canonical value $\z=0.0025$.  The rate is of course even larger for
greater values of $\z$.   We therefore see that the new contributions to the
low energy effective Hamiltonian from the $\221$ theory can lead to
significant deviations from the $b \to s \g$ predictions of the Standard
Model.

\newsec{Strong Interaction Corrections}

	QCD corrections to $b \to s \g$ decay have received considerable
attention during the past several years and are known to be very large in
the Standard Model \refs{\QCDcorrections,\Ciuchini}.  The analysis of
strong interaction effects upon
the rare radiative transition is most sensibly conducted within the
five-quark effective theory where large logarithms can be summed
using the renormalization group.   Since the structure
of the low energy effective theory does not sensitively depend upon
the precise nature of physics beyond the electroweak scale, the
computation of strong interaction corrections is similar in both the
$\221$ and $\twoone$ models.  We can therefore take over many
well-known results from prior $b \to s \g$ studies.

	We start by generalizing the effective Hamiltonian in \HeffI\ to
include operators that mix with the photon magnetic moment terms under the
action of QCD renormalization:
\eqn\Heff{H_\eff = - {4 \GF \over \sqrt{2}} {V^{ts}_\L}^* V^{tb}_\L
\sum_j C_j(\u) O_j(\u).}
We adopt the following conventional choice for the set of operators appearing
in the effective Hamiltonian:
\eqn\ops{\eqalign{
O_1 &= (\bar{s}_\a \g_\u P_- c^\b) (\bar{c}_\b \g^\u P_-
b^\a) \cr
O_2 &= (\bar{s}_\a \g_\u P_- c^\a) (\bar{c}_\b \g^\u P_-
b^\b) \cr
O_3 &= (\bar{s}_\a \g_\u P_- b^\a) \sum_q (\bar{q}_\b \g^\u P_- q^\b)
\cr
O_4 &= (\bar{s}_\a \g_\u P_- b^\b) \sum_q (\bar{q}_\b \g^\u P_- q^\a)
\cr
O_5 &= (\bar{s}_\a \g_\u P_- b^\a) \sum_q (\bar{q}_\b \g^\u P_+ q^\b)
\cr
O_6 &= (\bar{s}_\a \g_\u P_- b^\b) \sum_q (\bar{q}_\b \g^\u P_+ q^\a)
\cr
O_7 &= {e \over 16\pi^2} \mb \bar{s}_\a \o^{\u\v} P_+ b^\a F_{\u\v} \cr
O_8 &= {g_3 \over 16\pi^2} \mb \bar{s}_\a \o^{\u\v} P_+ (T^a)^\a_\b
b^\b G_{\u\v}^a. \cr}}
Here $\a$ and $\b$ represent color indices, while the summation over $q$
ranges over the five active quark flavors.  This list constitutes a
complete operator basis if the underlying full theory is the Standard,
two-Higgs doublet or minimal supersymmetric model.

	In the $\221$ effective theory however, new operators with different
chirality structures can arise.  In particular, we need to include the
four-quark terms
\eqn\newops{\eqalign{
O_9 &= \bigl({\mb\over\mc} \bigr) (\bar{s}_\a \g_\u P_- c^\b)
 (\bar{c}_\b \g^\u P_+ b^\a) \cr
O_{10} &= \bigl({\mb\over\mc} \bigr) (\bar{s}_\a \g_\u P_- c^\a)
 (\bar{c}_\b \g^\u P_+ b^\b) \cr}}
which are left-right analogues of $O_1$ and $O_2$.  The ratios of
the bottom and charm quark masses are incorporated into
their definitions to facilitate later mixing
computations involving these operators.  We also need to introduce
the flipped chirality partners $O'_1$ - $O'_{10}$ of $O_1$ - $O_{10}$
obtained by setting $P_\pm \to P_\mp$ in eqns.~\ops\ and \newops.
Most of these new operators will fortunately play no significant role in
our $b \to s \g$ analysis.  So the total number of operators that we
will actually need to consider is much smaller than 20.

	After performing a straightforward matching computation,
we find the following $W_1$ scale coefficient values in the limit of
vanishing up quark mass:
\foot{We retain the charm quark contributions to (4.4) even though they are
suppressed relative to the top quark terms by $\mc/\mt$.  This small factor
could in principle be offset by the ratio of KM angles in (4.5) in an
asymmetric left-right model.}
\eqn\coeffvalues{
\eqalign{
 C_2(\MW1) &= 1 \cr
 C_7(\MW1) &= F(x_t)+A^{tb} {\tilde F}(x_t) + A^{cb} \cr
 C_8(\MW1) &= G(x_t)+A^{tb} {\tilde G}(x_t) \cr
 C_{10}(\MW1) &= A^{cb} \cr}
\qquad\qquad
\eqalign{
 C'_2(\MW1) &=0 \cr
 C'_7(\MW1) &= (A^{ts})^* {\tilde F}(x_t)+(A^{cs})^* \cr
 C'_8(\MW1) &= (A^{ts})^* {\tilde G}(x_t) \cr
 C'_{10}(\MW1) &= (A^{cs})^* \cr}}
where
\eqn\Adefn{A^{\U\D} = \zg {m_\U \over m_b} {V_\R^{\U\D} \over
V_\L^{\U\D}} e^{i\a} \quad{\rm for}\quad U=u,c,t \quad{\rm and}\quad D=d,s,b.}
The functions $F$ and ${\tilde F}$ in the coefficients
of the photon magnetic moment operators $O^{(\prime)}_7$ were previously
specified in \Ffunctions.  The analogous functions for the gluon
magnetic moment operators $O^{(\prime)}_8$ are given by
\eqn\Gfunctions{\eqalign{
G(x) &= {x(2+5x-x^2) \over 8(x-1)^3} - {3x^2 \over 4(x-1)^4} \ln x \cr
{\tilde G}(x) &= -{4+x+x^2 \over 4(x-1)^2} + {3x \over 2 (x-1)^3} \ln x. \cr }}
All other operator coefficients vanish at the $W_1$ scale.

	The renormalization group mixing of the operators in our basis
set is governed by a $20 \times 20$ anomalous dimension matrix $\g$.
Since the strong interactions preserve chirality, the unprimed operators
in eqns.~\ops\ and \newops\ cannot mix with their primed counterparts
under the action of QCD.  Moreover, renormalization group mixing within
the two separate operator sectors is precisely the same.
Therefore, $\g$ decomposes into two identical $10 \times 10$
blocks.  The leading order structure of these blocks breaks up
into an $8 \times 8$ submatrix $\g_{8 \times 8}$ and a partially
overlapping $4 \times 4$ submatrix $\g_{4 \times 4}$.  The $8 \times 8$
matrix describes the mixing among $O^{(\prime)}_1$ - $O^{(\prime)}_8$ and
has been calculated by a number of groups \refs{\QCDcorrections,\Ciuchini}.
At this time, complete consensus regarding the exact values for all
the entries in $\g_{8 \times 8}$ has not been achieved.  While this lack of
agreement is disturbing, it is of relatively little practical importance
since all competing claims for $\g_{8 \times 8}$ yield nearly identical
numerical results for the $b \to s \g$ decay rates in the $\twoone$ and
$\221$ models.  We will use the recent results of Ciuchini {\it et al.} for
this matrix. The remaining $4 \times 4$ matrix overlaps
with $\g_{8 \times 8}$ and controls the mixing of the two new
four-quark operators in \newops\ into the dimension-five photon and
gluon magnetic moment operators.  Its entries can be extracted from the
computations of analogous mixings within $\g_{8 \times 8}$ and are
exhibited below:
\eqn\anomfourfour{\g_{4 \times 4} =
\bordermatrix{& O^{(\prime)}_7 & O^{(\prime)}_8 & O^{(\prime)}_9 &
 O^{(\prime)}_{10} \cr
O^{(\prime)}_7 & {16 / 3} & 0 & 0 & 0 \cr
O^{(\prime)}_8 & -{16 / 9} & {14 / 3} & 0 & 0 \cr
O^{(\prime)}_9 & {80 / 3} & -2 & -8 & 0 \cr
O^{(\prime)}_{10} & {32/  9} & 4/3 & -3 & 1\cr}{g_3^2 \over 8\pi^2} .}
All other entries in the $10 \times 10$ anomalous dimension blocks vanish.

	Once the anomalous dimension matrix is determined, it is
straightforward to solve the renormalization group equation which relates
coefficient values at $\mu=\MW1$ to those at $\mu=\mb$.  The solution
appears as
\eqn\newcoeffvalues{C_i(\mb) = \sum_{j,k} (S^{-1})_{ij} \>
\Bigl( \eta^{3\lambda_j/23} \Bigr) \> S_{jk} C_k(M_{W_1})}
where the $\lambda_j$'s in the exponent of $\n = \aS(M_{W_1})/\aS(\mb)$ are
the eigenvalues of ${\hat\g}=\g/(g_3^2/8\pi^2)$ and the rows of matrix
$S$ contain the corresponding eigenvectors.

	Assembling together the bottom scale coefficients and matrix elements
of all the operators in our basis set, we finally obtain the QCD corrected
$b \to s \g$ decay rate in the $\221$ model:
\eqn\bsphotrateII{
\Gamma(b \to s \g) = \displaystyle{{\GF^2 \mb^5 \over 32 \pi^4}} \aEM(\mb)
|{V_\L^{ts}}^* V_\L^{tb}|^2 \Bigl( |C_{7\eff}(\mb)|^2 +|C'_{7\eff}(\mb)|^2
\Bigr).}
The effective magnetic moment operator coefficients are given by
\eqn\cseveneff{\eqalign{
C_{7\eff}(\mb) &= C_{7\eff}(\mb)_{SM}
 + A^{tb} \Bigl[ \n^{16/23} {\tilde F}(x_t)
 + {8 \over 3} \bigl( \n^{14/23} - \n^{16/23} \bigr) {\tilde G}(x_t)\Bigr]
 + A^{cb} \sum_{i=1}^4 h'_i \n^{p'_i} \cr
C'_{7\eff}(\mb) &=
   (A^{ts})^* \Bigl[ \n^{16/23} {\tilde F}(x_t)
 + {8 \over 3} \bigl( \n^{14/23} - \n^{16/23} \bigr) {\tilde G}(x_t)\Bigr]
 + (A^{cs})^* \sum_{i=1}^4 h'_i \n^{p'_i} \cr}}
where
\eqn\cseveneffSM{C_{7\eff}(\mb)_{SM} = \n^{16/23} F(x_t)
 + {8 \over 3} \bigl( \n^{14/23} - \n^{16/23} \bigr) G(x_t)
 + \sum_{i=1}^8 h_i \n^{p_i}}
denotes the corresponding Standard Model result.  The coefficients $h_i$ and
powers $p_i$ entering into the last term have been discussed and tabulated
in the appendix of ref.~\MisiakII.  We simply quote them here
\eqn\hvalues{\eqalign{
(h_1,h_2,h_3,h_4,&h_5,h_6,h_7,h_8) = \cr
& (2.2996,-1.0880,-0.4286,-0.0714, -0.6494,-0.0380,-0.0186,-0.0057) \cr
(p_1,p_2,p_3,p_4,&p_5,p_6,p_7,p_8) = \cr
& (0.6087,0.6957,0.2609,-0.5217, 0.4086,-0.4230,-0.8994,0.1456) \cr}}
along with the $h'_i$ and $p'_i$ values
\eqn\hpvalues{\eqalign{
(h'_1,h'_2,h'_3,h'_4) &= \bigl( -0.6615,1.3142,0.0070,1.0070 \bigr) \cr
(p'_1,p'_2,p'_3,p'_4) &= \bigl( 0.6957,0.6087,-1.0435,0.1304 \bigr). \cr}}

	The radiative partial width in \bsphotrateII\ is regularization
and renormalization scheme independent as must be the case for any physical
observable.
\foot{We should point out that the coefficients $C^{(\prime)}_7$ and
$C^{(\prime)}_8$ in eqn.~\coeffvalues, the nonvanishing off-diagonal
$2 \times 2$ block in the anomalous dimension submatrix $\g_{4 \times 4}$
in eqn.~\anomfourfour, and the one-loop matrix elements of $O^{(\prime)}_9$
and $O^{(\prime)}_{10}$ are regularization scheme dependent.  These
quantities were all calculated in the fully anticommuting $\gfive$
dimensional regularization scheme.}
We normalize it to the QCD corrected generalization of the semileptonic rate
in \semirateI
\eqn\semirateII{
\Gamma(b \to c e \bar{\nu}_e) = {\GF^2 \mb^5 \over 192 \pi^3} |V^{cb}_\L|^2
\gmcmb \Bigl[ 1-{2 \over 3 \pi} \a_s(\mb) \fmcmb \Bigr]}
and form the ratio
\eqn\rateratioII{
R = {\Gamma(b \to s \g) \over \Gamma(b \to c e \bar{\nu}_e)}
 \simeq {\Gamma(\bar{B} \to X_s \g) \over \Gamma(\bar{B} \to X_c e
 \bar{\v}_e)}
 = {6 \over \pi} {\aEM(\mb)\over \displaystyle{\gmcmb}}
 {|C_7(\mb)_\eff|^2+|C'_7(\mb)_\eff|^2  \over
 1 - \displaystyle{2 \over 3 \pi} \a_s(\mb) \fmcmb}.}
The function $f$ appearing in these expressions encodes sizable
next-to-leading order strong interaction effects which we choose to
include and is numerically tabulated in ref.~\Cabibbo.
In order to restrict the parameter dependence of $R$ so that it can
be simply displayed, we will specialize to the left-right symmetric limit and
set $\gtwoL=\gtwoR$ and $|V_\L| = |V_\R|$.  $R$ then depends only upon the
three parameters $\z$, $\mt$ and $\a$.

	In \zetaplot, we plot $R$ as a function of the mixing angle $\z$
in both the $\221$ and $\twoone$ models with $\mt=150 \GeV$ and $\a=0$ held
fixed.
\foot{The graph in \zetaplot\ may be interpreted in the context of an
asymmetric left-right model by rescaling $\z \to \zg | V_\R / V_\L |$
provided $|A^{tb}| = |A^{ts}| >> |A^{cb}|, |A^{cs}|$.}
Comparing these curves with their QCD uncorrected counterparts in
\uncorrectedzetaplot, we see that the strong interactions
triple the $b \to s \g$ rate for very small values of $\z$.  The strong
interaction enhancement at larger values of $\z$ is less pronounced.
The reason behind this trend can be seen in the expressions for the
effective photon magnetic moment coefficients $C^{(\prime)}_{7\eff}(\mb)$
and $C_{7 \eff}(\mb)_{\S\M}$.  Recall that the disparity between these
coefficients stems mainly from the terms proportional to ${\tilde F}(x_t)$
in \cseveneff. This discrepancy is suppressed however by the QCD factor
$\n^{16/23}=0.67$.  The last term in \cseveneffSM\ overcomes this
suppression factor and leads to a net QCD enhancement of the $b \to s \g$
rate in both the $\221$ and $\twoone$ models.  But the strong interactions
tend to diminish the difference between these two theories' rates.

	The dependence of $R$ upon $\mt$ for $\z=0.0025$ and $\a=0$ is
illustrated in \mtplot.  Both the $\221$ theory and Standard Model results
grow with increasing top mass.  For all $\mt$ above the present
experimental lower bound of $113 \GeV$ \CDF, we see that
that the former is greater than the latter by at least $30 \%$ for this
choice of parameters.  Such a variation is potentially large enough to
differentiate between these two models given current theoretical and
future experimental uncertainties.   Other regions in parameter space can of
course yield larger or smaller discrepancies.   We believe however that the
results in \mtplot\ are representative for most left-right symmetric
models.

	Finally, we plot $R$ as a function of the phase angle $\a$ with
$\z=0.0025$ and $\mt=150 \GeV$ held fixed in \alphaplot.  As can be seen in
the graph, maximum constructive and destructive interference between the
Standard Model and $\221$ contributions to the $b \to s \g$ effective
Hamiltonian occur for $\a=0$ and $\a=\pi$ respectively.  Distinguishing
between the two theories is consequently easiest for values of $\a$ near
these two endpoints.  Such values are fortunately favored in the $\221$ model
to avoid excessive CP violation \LangackerII.

	In conclusion, we have analyzed the rare $b \to s \g$ decay mode
in $\221$ extensions of the Standard Model.  We have found that mixing
between left and right $W$ bosons in such models can lead to sizable
new contributions to the effective Hamiltonian for this radiative process
even though the mixing angle $\z$ is constrained to be quite small.  QCD
corrections diminish the disparity between the $b \to s \g$ rates in
the $\221$ and $\twoone$ theories.  However for reasonable ranges of parameter
values, the decay rates can be distinguished and used to probe for new
physics beyond the Standard Model.

\bigskip
\centerline{\bf Acknowledgements}
\bigskip

	We thank Andrzej Buras and Mark Wise for helpful discussions.
PC thanks the Aspen Center for Physics and the high energy theory group at
Technische Universit\"at M\"unchen where work on this paper was performed for
their warm hospitality.

\vfill\eject

\listrefs
\listfigs
\bye